\documentstyle[12pt]{article}
\def\one{1\hskip-.37em 1}
\def\ir{{\rm I}\hskip-.2em{\rm R}}
\def\half{\textstyle{\frac{1}{2}}}

\def\iH{{\rm I}\hskip-.2em{\rm H}}
\def\ra{\rightarrow}
\def\tint{{\textstyle\int}}
\def\d{\partial}
\def\o{\overline}
\def\b{\begin{eqnarray*}}     
\def\e{\end{eqnarray*}}       
\def\bn{\begin{eqnarray}}     
\def\en{\end{eqnarray}}       
\def\<{\langle}
\def\>{\rangle}

\def\{{\lbrace}
\def\}{\rbrace}
\title{Geometric Quantization from a \\Coherent State Viewpoint}
\author{John R. Klauder\\
Departments of Physics and Mathematics\\
University of Florida\\
Gainesville, Fl  32611}
\begin{document}
\maketitle
\begin{abstract}
A fully geometric procedure of quantization that utilizes a natural and
necessary metric on phase space is reviewed and briefly related to the goals
of the program of geometric quantization.
\end{abstract}
\section*{Introduction and Background}
\subsubsection*{Purpose and achievements}
The goal of the present work is to present a {\it conceptually simple,
geometric prescription for quantization}. Such a goal has been, and
continues to be, the subject of a number of research efforts. In the present
paper we shall look at the problems faced by this program from the point of
view of coherent states. It will be our conclusion that the symplectic
geometry of classical mechanics, augmented by a natural, and even necessary,
metric on the classical phase space, are the only essential ingredients to
provide a quantization scheme that is fully geometric in character and one
that can be expressed in a coordinate-free form. In a surprising sense we
shall see that coherent states---far from being optional---are in fact an
automatic consequence of this process of quantization. It is fair to say at
the outset that we will have very little to say about the {\it methods} used
in the program of geometric quantization; instead, what we adopt is the {\it
goal} of this program. As will be clear our methods, especially for
dynamical questions, are quite different. Moreover, we shall succeed in
correctly quantizing a large number (a dense set) of Hamiltonians. In that
sense the procedures to be described offer a fully satisfactory solution to
the program of geometric quantization, namely to provide an intrinsic and
coordinate-free prescription for quantization which agrees with known
quantum mechanical results.

Let us start our story with a brief overview of just what it is about
classical and quantum mechanics that gives rise in the first place to the
problem that the program of geometric quantization addresses. For ease of
exposition we shall assume that our phase space admits global coordinates
unless explicitly stated otherwise. And for pedagogical reasons we shall
focus  on just a single degree of freedom; if we don't understand one degree
of freedom well, then we probably don't understand much else!
\subsubsection*{Symplectic manifolds and classical mechanics}
For a single degree of freedom, phase space is a two-dimensional manifold
$M$ equipped with a {\it symplectic form} $\omega$, which is a closed,
nondegenerate two form. As a closed form it is locally exact, that is
$\omega=d\theta$, where $\theta$ is a one form and $d$ denotes the operation
of exterior derivative. This one form is not unique and it follows from the
fact that $d^2=0$ that we can replace $\theta$ by $\theta+dG$, where  $G$ is
a zero form or scalar function. For a general $G$ it follows that
  $$d(\theta+dG)=d\theta+d^2G=d\theta=\omega\;.$$
In addition we introduce a scalar function $H$ on the manifold $M$ which
takes on a unique value at each point $x\in M$. Now consider a path $x(t)$,
$0\equiv t'\leq t\leq t''\equiv T$, in the manifold $M$ parameterized by the
variable $t$ (``time''), so that $\theta$, $G$, and $H$ all become functions
of $t$ through their dependence on the phase-space points along the path.
Thus we can construct an action integral according to
  $$    I=\int(\theta+dG-H\,dt)=\int
(\theta-H\,dt)+G''-G'\;,$$
and extremal variations, which hold both end points $x(t')$ and $x(t'')$
fixed, lead to the sought for classical trajectory through the manifold. The
associated equations of motion do not involve $G$. In that sense there is an
{\it equivalence class} of actions all of which lead to the same classical
equations of motion and thus the same classical trajectory.

This scenario may be given a more familiar look with the introduction of
canonical coordinates. The Darboux theorem assures us that charts of local
canonical coordinates $p$ and $q$ exist for which $\theta=p\,dq$,
$\omega=dp\wedge dq$, $G=G(p,q)$, and $H=H(p,q)$. In these terms
$$I=\int[p\,dq+dG(p,q)-H(p,q)\,dt]=\int[p\,dq-H(p,q)\,dt]+G(p'',q'')
- -G(p',q')\,,$$
where $p',q'\equiv p(0),q(0)$, etc. Insisting that the vanishing of the
first-order variation holding the end points $p'',q''$ and $p',q'$ fixed
characterizes the classical trajectory and leads to the usual Hamiltonian
equations of motion, namely
$${\dot q}=\frac{\d H(p,q)}{\d p}\;,\;\;\;\;\;\;{\dot p}=-\frac{\d H(p,q)}
{\d q}\;.$$
These equations of motion are independent of $G$, and thus we have an
equivalence class of actions all of which lead to the same classical
equations of motion.

New canonical coordinates, say $\o p$ and $\o q$, are invariably related to
the original canonical coordinates through the one form
  $${\o p}\,d{\o q}=p\,dq+dF({\o q},q)$$
for some function $F$.
It follows, therefore, that for a general $G$ and $F$, one may always find a
function $\o G$ such that
  $$I=\tint[{\o p}\,d{\o q}+d{\o G}({\o p},{\o q})-{\o H}({\o p},{\o
q})\,dt]=\tint[{\o p}\,d{\o q}-{\o H}({\o p},{\o q})\,dt]+{\o G}({\o
p''},{\o q''})-{\o G}({\o p'},{\o q'})\,,$$
where ${\o H}({\o p},{\o q})=H(p,q)$.
Extremal variation of this version of the action holding the end points
fixed leads to Hamilton's equations expressed in the form
  $${\dot{\o q}}=\frac{\d{\o H}({\o p},{\o q})}{\d {\o
p}}\;,\;\;\;\;\;\;{\dot{\o p}}=-\frac{\d{\o H}({\o p},{\o q})}{\d {\o
q}}\;.$$
This form invariance of Hamilton's equations among canonical coordinates is
what distinguishes this family of coordinate systems in the first place.
Such form invariance is the clue that after all there is something of a
geometrical nature underlying this structure, and that geometry is in fact
the symplectic geometry briefly discussed above.

It is noteworthy that a given Hamiltonian expressed, say, by
$H(p,q)=\frac{1}{2}(p^2+q^2)+q^4$, in one set of canonical coordinates,
could, in a new set of canonical coordinates, be expressed simply as ${\o
H}({\o p},{\o q})={\o p}$. If we interpret the first expression as
corresponding physically to an anharmonic oscillator, a natural question
that arises is how is one to ``read'' out of the second expression that one
is dealing with an anharmonic oscillator. This important question will
figure significantly in our study!
\subsubsection*{Old quantum theory}
In the old quantum theory one approximately quantized energy levels by the
Bohr-Sommerfeld quantization scheme in which
  $$\oint p\,dq=(n+{\half})2\pi\hbar\;,$$
where the integral corresponds to a closed contour in phase space at a
constant energy value. This integral is to be taken in some canonical
coordinate system, but which one? Since two canonical coordinate systems are
related by
  $${\o p}\,d{\o q}=p\,dq+dF({\o q},q)\;,$$
it follows when $M$ is simply connected that
  $$\oint{\o p}\,d{\o q}=\oint p\,dq\;,$$
and so which canonical coordinates are used doesn't matter---they all give
the same result. In coordinate-free language this remark holds simply
because
  $$\oint p\,dq=\int dp\wedge dq=\int d{\o p}\wedge d{\o
q}\equiv\int\omega\;.$$
Thus we also learn one answer to the question posed above, namely the
physics of the mathematical expression for the Hamiltonian given simply by
$\o p$ is coded into the orbits and into the coordinate-invariant
phase-space areas captured by $\int\omega$ for each value of the energy.
\subsubsection*{New quantum theory}
The royal route to quantization, according to Schr\"odinger for example,
consists first of introducing a Hilbert space of functions $\psi(x)$,
defined for $x\in\ir$, each of which satisfies $\int|\psi(x)|^2\,dx<\infty$.
Next the classical phase-space variables $p$ and $q$ are ``promoted'' to
operators, $p\ra -i\hbar\d/\d x$ and $q\ra x$, which act by differentiation
and multiplication, respectively. More general quantities, such as the
Hamiltonian, become operators according to the rule
  $$H(p,q)\ra{\cal H}=H(-i\hbar\d/\d x,x)\;,$$
an expression that may have ordering ambiguities but which we will ignore in
favor of the deeper question: In which canonical coordinate systems does
such a quantization procedure work? Dirac's answer is: ``This assumption is
found in practice to be successful only when applied with the dynamical
coordinates and momenta referring to a Cartesian system of axes and not to
more general curvilinear coordinates.''\cite{dir}. In other words, the
correctness of the Schr\"odinger rule of quantization depends on using the
right coordinates, namely Cartesian coordinates. It is worth emphasizing
that Cartesian coordinates can only exist on a {\it flat space}. Likewise in
the prescription of Heisenberg which asks (among other things) that the
classical canonical variables $p$ and $q$ be replaced by operators $P$ and
$Q$ that satisfy the commutation relation $[Q,P]=i\hbar$; this rule must
also be applied only in Cartesian coordinates.

An analogous feature is evident in the Feynman phase-space path integral
formally given by
  $${\cal M}\int e^{(i/\hbar)\int_0^T[p{\dot q}-H(p,q)]\,dt}\,{\cal
D}p\,{\cal D}q\;.$$
 Despite the appearance of this expression as being covariant under a change
of canonical coordinates, it is, as commonly known, effectively undefined.
One common way to define this expression is by means of a
lattice formulation one form of which is given by
  $$\lim_{N\ra\infty}M\int\exp\{{\textstyle\frac{i}{\hbar}}\Sigma_0^N[
p_{l+\frac{1}{2}}(q_{l+1}-q_l)-\epsilon
H(p_{l+\frac{1}{2}},{\textstyle\frac{1}{2}}(q_{l+1}+q_l))]\}\,\Pi_0^N\,dp_{l
+\frac{1}{2}}\,\Pi_1^N\,dq_l$$
where $\epsilon\equiv T/(N+1)$, $q_{N+1}\equiv q''$, $q_0\equiv q'$, and
$M=(2\pi\hbar)^{-(N+1)}$ is a suitable normalization factor. The indicated
limit is known to exist for a large class of Hamiltonians leading to
perfectly acceptable (Weyl-ordered) quantizations. However, it is clear that
unlike the formal continuum path integral, this lattice formulation is {\it
not} covariant under canonical coordinate transformations; in other words,
this lattice expression will lead to the correct quantum mechanics only in a
limited set of canonical coordinates, namely the Cartesian set mentioned by
Dirac.

This then is the dilemma that confronts us. The ``new'' quantization of
Schr\"odinger, Heisenberg, and Feynman---the {\it correct} quantization from
all experimental evidence---seems to depend on the choice of coordinates.
This is clearly an unsettling state of affairs since nothing physical, like
quantization, should depend on something so arbitrary as the choice of
coordinates!
\subsubsection*{Geometric quantization}
There are two attitudes that may be taken toward this apparent dependence of
the very act of quantization on the choice of coordinates. The first view
would be to acknowledge the ``Cartesian character'' that is seemingly part
of the procedure. The second view would be to regard it as provisional and
seeks to find a quantization formulation that eliminates this apparently
unphysical feature of the current approaches. There is much to be said for
this second view. After all Newton's equations for particle dynamics
expressed originally in Cartesian coordinates may be given a tensorial
formulation that is valid in all coordinate systems. There seems to be no
apparent reason that some similar reformulation of the usual quantization
procedures may not do the same for quantum mechanics.

The goal of eliminating the dependence on Cartesian coordinates in the
standard approaches is no doubt one of the motivations for several programs
such as geometric quantization \cite{syn}, deformation quantization
\cite{fla}, etc. In the first of these programs, for example, one finds the
basic ingredients: (i) prequantization and (ii) polarization (real and
complex), which define the framework, i.e, the {\it kinematics}, and (iii)
one of several proposals to deal with {\it dynamics}. Despite noble efforts,
it is not unfair to say that to date only a very limited class of dynamical
systems can be treated in the geometric quantization program which also
conform with the results of quantum mechanics.

The approach that we shall adopt takes the other point of view seriously,
namely that the ``Cartesian character'' is not to be ignored. As we shall
see when this feature is properly understood and incorporated, {\it a
genuine geometric interpretation of quantization can be rigorously developed
that agrees with the predictions of Schr\"odinger, Heisenberg, and Feynman,
and does so for a wide (dense) set of Hamiltonians}.
\section*{Coherent States}
The concept of coherent states is sufficiently broad by now that there are
several definitions. Our definition is really an old one \cite{k63}, and a
very general one at that, so general that it captures essentially all other
definitions. We start with a label space ${\cal L}$, which may often be
identified with the classical phase space $M\!$, and a continuous map from
points in the label space to (nonzero) vectors in a Hilbert space $\iH$ (see
below). For concreteness let us choose the label space as the phase space
for a single degree of freedom. Then each point in $M$ may be labelled by
canonical coordinates $(p,q)$, and we use that very set to identify the
coherent state vector itself: $|p,q\>$ or $\Phi[p,q]$. If we choose a
different set of canonical coordinates, say $({\o p},{\o q})$ to identify
the same point in $M$, then we associate the new coordinates to the {\it
same} vector $|{\o p},{\o q}\>\equiv|p,q\>$, or even better ${\o\Phi}[{\o
p},{\o q}]\equiv\Phi[p,q]$.  Although it is not necessary to do so, we shall
specialize to coherent states that are unit vectors,
$\<p,q|p,q\>=1=(\Phi[p,q],\Phi[p,q])$, for all points $(p,q)\in M$.
We place only two requirements on this map from $M$ into $\iH$:

(1) {\it continuity}, which can be stated as joint continuity in both
arguments of the coherent state overlap ${\cal
K}(p'',q'';p',q')\equiv\<p'',q''|p',q'\>$; and

(2) {\it resolution of unity}, for which a positive measure $\mu$ exists
such that
  $$\one\equiv\int|p,q\>\<p,q|\,d\mu(p,q)\;,$$
where $\one$ is the unit operator. This last equation may be understood as
  \b \<\phi|\psi\>\!\!\!&=&\!\!\!\!\int\<\phi|p,q\>\<p,q|\psi\>\,
d\mu(p,q)\;,\\
  \<p'',q''|\psi\>\equiv\psi(p'',q'')\!\!\!&=&\!\!\!\!\int{\cal K}(p'',q''
;p,q)\,\psi(p,q)\,d\mu(p,q)\;,\\
  {\cal K}(p'',q'';p',q')\!\!\!&=&\!\!\!\!\int{\cal K}(p'',q'';p,q)\, {\cal
K}(p,q;p',q')\,d\mu(p,q)\;.  \e
Each successive relation has been obtained from the previous one by
specialization of the vectors involved. The last relation, in conjunction
with the fact that ${\cal K}(p'',q'';p',q')^*={\cal K}(p',q';p'',q'')$,
implies that $\cal K$ is the kernel of a projection operator onto a proper
subspace of $L^2(\ir^2,d\mu)$ composed of bounded, continuous functions that
comprise the Hilbert space of interest. The fact that the representatives
are bounded follows from our choice of coherent states that are all unit
vectors. Although no group need be involved in our definition of coherent
states, it is evident that when a group is present simplifications may
occur. This is true for the canonical coherent states to which we now
specialize.

With $Q$ and $P$ self adjoint and irreducible and $[Q,P]=i\hbar$, the
canonical coherent states defined with help of the unitary Weyl group
operators for all $(p,q)\in\ir^2\equiv M$ by
$$|p,q\>=e^{-iqP/\hbar}\,e^{ipQ/\hbar}\,|\eta\>\;,\;\;\;\;\;\<\eta|\eta\>=1\
;$$
are a standard example for which $d\mu(p,q)=dp\,dq/2\pi\hbar$. The
resolution of unity holds in this case for {\it any} normalized fiducial
vector $|\eta\>$. However, a useful specialization occurs if we insist that
$(\Omega Q+iP)|\eta\>=0$, $\Omega>0$, leading to the ground state of an
harmonic oscillator. In that case
$$\<p',q'|p,q\>=\exp\{(i/2\hbar)(p'+p)(q'-q)-(1/4\hbar)[\Omega^{-1}(p'-p)^2+
\Omega(q'-q)^2]\}\;.$$
We note in passing that the more usual resolutions of unity may be obtained
as limits. In particular,  \b
\int\lim_{\Omega\ra\infty}|p,q\>\<p,q|\,dp\,dq/2\pi\hbar=\int|q\>\<q|\,dq=
\one\;,\\ \int\lim_{\Omega\ra
0}|p,q\>\<p,q|\,dp\,dq/2\pi\hbar=\int|p\>\<p|\,dp=\one\;, \e  where the
formal vectors $|q\>$ satisfy $Q|q\>=q|q\>$and $\<q'|q\>=\delta(q'-q)$, and
correspondingly for $|p\>$.

The normalized canonical coherent states $|p,q\>$ that follow from the
condition $(\Omega Q+iP)|\eta\>=0$ are in fact analytic functions of the
combination $\Omega q+ip$ apart from a common prefactor. When that prefactor
is removed from the vectors and put into the integration measure, one is led
directly to the Segal-Bargmann representation by holomorphic functions.
\subsubsection*{Symbols}
Generally, and with the notation $\<(\cdot)\>\equiv\<\eta|(\cdot)|\eta\>$,
it follows from the commutation relations that $\<p,q|P|p,q\>=p+\<P\>$ and
$\<p,q|Q|p,q\>=q+\<Q\>$; if
$\<p,q|P|p,q\>=p$ and $\<p,q|Q|p,q\>=q$ we say that the fiducial vector is
physically centered. Observe that the labels of the coherent state vectors
are {\it not eigenvalues} but {\it expectation values}; thus there is no
contradiction in specifying both $p$ and $q$ simultaneously.

For a general operator ${\cal H}(P,Q)$ we introduce the {\it upper} symbol
 \b H(p,q)\!\!\!&\equiv&\!\!\!\<p,q|{\cal H}(P,Q)|p,q\>\\
&=&\!\!\!\<\eta|{\cal H}(P+p,Q+q)|\eta\>\\
   &=&\!\!\!{\cal H}(p,q)+{\cal O}(\hbar;p,q)\;,  \e
and, when it exists, the {\it lower} symbol $h(p,q)$ implicitly defined
through the relation
  $${\cal H}=\int h(p,q)\,|p,q\>\<p,q|\,dp\,dq/2\pi\hbar\;. $$
We note that for the harmonic oscillator fiducial vector lower symbols exist
for a dense set of operators, and generally $H(p,q)-h(p,q)\simeq O(\hbar)$.
The association of an operator $\cal H$ with the function $h(p,q)$ is an
example of what goes under the name of Toeplitz quantization today
\cite{jaf}.
\subsubsection*{Differentials}
Several differential expressions are already implicitly contained within the
coherent states. The first is the canonical one form
  $$\theta\equiv
i\hbar\<\;|d|\;\>=i\hbar(\Phi,d\Phi)=i\hbar\Sigma_n\phi^*_n\,d\phi_n$$
written in coordinate-free notation, or alternatively,
  $$\theta=i\hbar\<p,q|d|p,q\>=p\,dq+\<P\>\,dq-\<Q\>\,dp=p\,dq $$
using canonical coordinates, and where we have ended with a physically
centered fiducial vector. In coordinate-free notation it follows that
  $$\omega\equiv d\theta =i\hbar\Sigma_n d\phi^*_n\wedge d\phi_n\;,$$
and in canonical coordinates that
  $$\omega=dp\wedge dq=d{\o p}\wedge d{\o q}\;,$$
along with $d\omega=0$ which follows directly. A useful Riemannian metric
is given first in coordinate-free notation by
  \b  d\sigma^2\!\!\!&\equiv&\!\!\! 2\hbar^2[
 |\!|d|\;\>|\!|^2-|\<\;|d|\;\>|^2]\\
&=&\!\!\!2\hbar^2\Sigma_{n,m}\,d\phi^*_n(\delta_{nm}-\phi_n\phi^*_m)d\phi_m\
;,   \e
and second in canonical coordinates by
  \b d\sigma^2(p,q)\!\!\!&=&\!\!\! \hbar(dp^2+dq^2)\;,\;\;\;\;\;\;\;\;\;
(\Omega=1)\;,\\
  d\sigma^2({\o p},{\o q})\!\!\!&=&\!\!\!\hbar[A({\o p},{\o q})d{\o
p}^2+B({\o p},{\o q})d{\o p}\,d{\o q}+C({\o p},{\o q})d{\o q}^2]\;.  \e
In the next to the last line the line element is expressed in the Cartesian
form it takes for a Gaussian fiducial vector, while in the last line is the
expression of the flat metric in general canonical coordinates.
\subsubsection*{Canonical and unitary transformations}
In classical mechanics canonical transformations may either be viewed as
passive or active. Passive transformations leave the point in phase space
fixed but change the coordinates by which it is described; active
transformations describe a flow of points in phase space against a fixed
coordinate system. The best known example of an active transformation is the
continuous unfolding in time of a dynamical evolution. In quantum mechanics
unitary transformations are presumed to play the role that canonical
transformations play in the classical theory \cite{shi}. If $p\ra P$ and
$q\ra Q$, then it follows that ${\o
p}=(p+q)/\sqrt{2}\ra(P+Q)/\sqrt{2}\equiv{\o P}$ and ${\o
q}=(q-p)/\sqrt{2}\ra(Q-P)/\sqrt{2}\equiv{\o Q}$, and moreover there exists a
unitary operator $U$ such that ${\o P}=U^\dagger PU$ and ${\o Q}=U^\dagger
QU$. Consider instead the classical canonical transformation ${\tilde
p}\equiv(p^2+q^2)/2\ra{\tilde P}$ and ${\tilde
q}\equiv\tan^{-1}(q/p)\ra{\tilde Q}$. As basically a transformation to polar
coordinates this canonical transformation is well defined except at the
single point $p=q=0$. However, the associated quantum operators in this case
cannot be connected by a unitary transformation to the original operators
$P$ and $Q$ (because ${\tilde P}\geq0$ and the spectrum of an operator is
preserved under a unitary transformation). Thus some passive canonical
transformations have images in unitary transformations while others
definitely do not.

Using coherent states it is possible to completely disconnect canonical
transformations and unitary transformations. Consider the transformations of
the upper and lower symbols in the following example:
  \b
{\half}(p^2+q^2)\!\!\!&=&\!\!\!\<p,q|\,{\half}(P^2+Q^2-\hbar)|p,q\>\\&=&\!\!
\!\<{\tilde p},{\tilde q}|\,{\half}(P^2+Q^2-\hbar)|{\tilde p},{\tilde
q}\>={\tilde p}\;,\\
{\half}(P^2+Q^2+\hbar)\!\!\!&=&\!\!\!\int{\half}(p^2+q^2)\,|p,q\>\<p,q|\,dp\
,dq/2\pi\hbar\\&=&\!\!\!\int{\tilde p}\,|{\tilde p},{\tilde q}\>\<{\tilde
p},{\tilde q}|\,d{\tilde p}\,d{\tilde q}/2\pi\hbar\;.  \e
Observe in this example how the operators and coherent state vectors have
remained completely fixed as the coordinates have passed from $(p,q)$ to
$({\tilde p},{\tilde q})$. Of course, one may also introduce separate and
arbitrary unitary transformations of the operators and vectors, e.g. $P\ra
VPV^\dagger$, and $|p,q\>\ra V|p,q\>$, etc., which have the property of
preserving inner products.
\section*{Shadow Metric and Cartesian Coordinates}
The form of the metric $d\sigma^2$ was given earlier for a special (harmonic
oscillator) fiducial vector. If instead we consider a general fiducial
vector $|\eta\>$, then it follows that
  $$d\sigma^2=\<(\Delta Q)^2\>\,dp^2+\<\Delta P\Delta Q +\Delta Q\Delta
P\>\,dp\,dq+\<(\Delta P)^2\>\,dq^2\;,$$
which shows itself to be always {\it flat}; thus this is a property of the
{\it Weyl group} and not of the fiducial vector. Here $\Delta P\equiv
P-\<P\>$,  etc. Unlike the symplectic form or the Hamiltonian, for example,
the metric is typically $O(\hbar)$ and thus it is essentially nonclassical.

Indeed, any quantization scheme in which the Weyl operators and Hilbert
space vectors appear leads to the metric $d\sigma^2$, whether it is
intentional or not. Such schemes may not {\it use} the metric, but it is
nevertheless there.

We assert that physics resides in Cartesian coordinates, and more
particularly in the coordinate form of the metric. Suppose
$d\sigma^2=\hbar(dp^2+dq^2)$, then it follows that
$H(p,q)=\frac{1}{2}(p^2+q^2)$ implies, as before, that ${\cal
H}=\frac{1}{2}(P^2+Q^2-\hbar)$. On the other hand, if instead
$d\sigma^2=\hbar[(2{\tilde p})^{-1}d{\tilde p}^2+(2{\tilde p})d{\tilde
q}^2]$, then ${\tilde H}({\tilde p},{\tilde q})={\tilde p}$ implies that
${\cal H}=\frac{1}{2}(P^2+Q^2-\hbar)$. In other words, {\it the physical
meaning of the coordinatized mathematical expression for some classical
quantity is coded into the coordinate form of the metric!} This remark is
already true at the classical level, namely one needs a ``shadow'' flat
metric on the classical phase space, or at least on a copy of it, so that
one can ascribe physical meaning to the coordinatized mathematical
expressions for one or another classical quantity. If the flat shadow metric
is expressed in Cartesian coordinates, then one may interpret an expression
such as $\frac{1}{2}(p^2+q^2)+q^4$ as truly representing a physical, quartic
anharmonic oscillator; if the flat shadow metric is {\it not} expressed in
Cartesian coordinates, then no such physical interpretation of such a
mathematical expression is justified.

Although we have originally introduced the phase-space metric in the quantum
theory via its construction in terms of coherent states, we now see that we
can alternatively view the phase-space metric (modulo a coefficient $\hbar$)
as an auxiliary classical expression that provides physical meaning for
coordinatized expressions of the classical theory.
\section*{Quantization and Continuous-time \\Regularization}
It should be self evident that quantization relates to physical systems
inasmuch as the quantization of a particular Hamiltonian is designed to
generate the spectrum appropriate to that physical system. Consider again
the formal phase-space path integral given by
  $${\cal M}\int e^{(i/\hbar)\int_0^T[p{\dot q}+{\dot
G}(p,q)-h(p,q)]\,dt}\,{\cal D}p\,{\cal D}q\;.$$
We have already stressed that this expression is not mathematically defined,
and now we emphasize that in fact it has no physics as well because there is
no way of telling to which physical system the coordinatized expression for
the Hamiltonian corresponds. In short, the formal phase-space path integral
expression has neither mathematical nor physical meaning as it stands!

We will remedy this situation in a moment, but there is one ``toy'' analog
worth introducing initially. Consider the conditionally convergent integral
that is given a definition through the introduction of a regularization and
its removal as in the expression
  $$\int_{-\infty}^\infty
e^{iy^2/2}\,dy\equiv\lim_{\nu\ra\infty}\int_{-\infty}^{\infty}e^{iy^2/2-y^2/
2\nu}\,dy=\sqrt{2\pi i}\;.$$
Other regularizations may lead to the same answer, or in fact they may lead
to different results; the physical situation should be invoked to choose the
relevant one.

Now let us introduce a continuous-time convergence factor into the formal
phase-space path integral in the form
 \b\lim_{\nu\ra\infty}\!\!\!\!\!\!&&\!\!\!\!\!\!{\cal M}_\nu\int
e^{(i/\hbar)\int_0^T[p{\dot q}+{\dot
G}(p,q)-h(p,q)]\,dt}\,e^{-(1/2\nu)\int_0^T({\dot p}^2+{\dot
q}^2)\,dt}\,{\cal D}p\,{\cal D}q\\
 &=&\!\!\lim_{\nu\ra\infty}2\pi\hbar\,e^{\hbar\nu T/2}\int
e^{(i/\hbar)\int_0^T[p\,dq+dG(p,q)-h(p,q)\,dt]}\,d\mu_W^\nu(p,q)\;.  \e
In the first line we have formally stated the form of the regularization,
while in the second line appears the proper mathematical statement it
assumes after some minor rearrangement. The measure $\mu_W^\nu$ is a
two-dimensional Wiener measure expressed in Cartesian coordinates on the
plane as signified by the metric $dp^2+dq^2$ that appears in the first line
in the regularization factor. {\it Here enters the very shadow metric
itself, used to give physical meaning to the coordinatized form of the
Hamiltonian, and which now additionally underpins a rigorous regularization
for the path integral!} As Brownian motion paths, with diffusion constant
$\nu$, almost all paths are continuous but nowhere differentiable. Thus the
initial term $\int p\,dq$ needs to be defined as a {\it stochastic integral}
and we choose to do so in the Stratonovich form, namely as $\lim
\Sigma\frac{1}{2}(p_{l+1}+p_l)(q_{l+1}-q_l)$, where $q_l\equiv
q(l\epsilon)$, etc., and the limit refers to $\epsilon\ra0$ \cite{str}. This
prescription is generally different from that of It\^o, namely $\lim \Sigma
p_l(q_{l+1}-q_l)$, due to the unbounded variation of the Wiener paths
involved. Observe, in the second line above, for each $0<\nu<\infty$, that
{\it no mathematical ambiguities remain}, i.e., the expression is completely
well defined. As we note below not only does the limit exist but it also
provides the correct solution to the Schr\"odinger equation  for a dense set
of Hamiltonian operators.

The continuous-time regularization involved, or its Wiener measure
counterpart, involves pinning the paths $p(t),q(t)$ at $t=T$ and at $t=0$ so
that $(p'',q'')=(p(T),q(T))$ and $(p',q')=(p(0),q(0))$. This leads to an
expression of the form $K(p'',q'',T;p',q',0)$, which may be shown to be
  \b K(p'',q'',T;p',q',0)\!\!\!&\equiv&\!\!\!\<p'',q''|\,e^{-i{\cal
H}T/\hbar}|p',q'\>\;,\\
    |p,q\>\!\!\!&\equiv&\!\!\!
e^{-iG(p,q)/\hbar}\,e^{-iqP/\hbar}\,e^{ipQ/\hbar}\,|\eta\>\;,\;\;\;\;(Q+iP)|
\eta\>=0\;,\\
   {\cal H}\!\!\!&\equiv&\!\!\!\int h(p,q)|p,q\>\<p,q|\,dp\,dq/2\pi\hbar\;.
 \e
In brief, the regularization chosen {\it automatically} leads to a coherent
state representation, and, in addition, it {\it selects} the Hamiltonian
operator determined by the lower symbol. There are three technical
requirements for this representation to hold \cite{dkl}:
\b  (1)&&\;\;\;\;\;\int h^2(p,q)\,e^{-A(p^2+q^2)}\,dp\,dq<\infty\;,\;\;\;\;
{\rm for \; all}\;\;A>0\;,\\
   (2)&&\;\;\;\;\;\int h^4(p,q)\,e^{-B(p^2+q^2)}\,dp\,dq<\infty\;,\;\;\;\;
{\rm for\; some}\;\; B<{\half}\;,\\
  (3)&&\;\;\;\;\;{\cal H}\;\; {\rm \;is\; e. s. a.\; on
}\;\;D=\{\Sigma_0^Na_n|n\>:\;a_n\in{\bf C},\;N<\infty\}\;,  \e
where the orthonormal states
$|n\>\equiv(1/\sqrt{n!})[(Q-iP)/\sqrt{2\hbar}]^n|\eta\>$, $n\geq0$.
Thus this representation includes (but is not limited to) {\it all
Hamiltonians that are Hermitian, semibounded polynomials of the basic
operators $P$ and $Q$}. We note that $G$ generally serves as an unimportant
gauge; however, if the topology of $M$ is not simply connected then $G$
contains the Aharanov-Bohm phase \cite{ree}. Observe that the propagator
formula also has an {\it analog physical system}, namely a two-dimensional
particle moving on a flat plane in the presence of a constant magnetic field
perpendicular to the plane. The limit in which the mass of the particle goes
to zero projects the system onto the first Landau level.

The point of using the Stratonovich prescription for stochastic integrals is
that the ordinary rules of calculus still apply \cite{str}. Thus the rule
for a canonical transformation given earlier, namely ${\o p}\,d{\o
q}=p\,dq+dF({\o q},q)$, still applies to Brownian motion paths.
Consequently, just as in the classical case a function ${\o G}({\o p},{\o
q})$ exists so that after such a canonical coordinate transformation
 \b {\o K}({\o p}'',{\o q}'',T;{\o p}',{\o q}',0)\!\!\!\!&=&\!\!\!\!\<{\o
p}'',{\o q}''|\,e^{-i{\cal H}T/\hbar}\,|{\o p}',{\o q}'\>\\
  \!\!\!\!&=&\!\!\!\!\lim_{\nu\ra\infty}2\pi\hbar\, e^{\hbar\nu T/2}\int
e^{(i/\hbar)\int_0^T[{\o p}d{\o q}+d{\o G}({\o p},{\o q})-{\o h}({\o p},{\o
q})dt]}\,d{\o\mu}_W^\nu({\o p},{\o q})\;. \e
Here ${\o\mu}_W^\nu$ denotes two-dimensional Wiener measure on the flat
plane expressed in general canonical coordinates.
\subsubsection*{Coordinate-free formulation}
The covariant transformation of the propagator indicated above implies that
a coordinate-free representation exists. We first introduce Brownian motion
as a map $\rho(t;0):\,M\times M\ra\ir^+$, $t>0$, with
$\lim_{t\ra0}\rho=\delta$, $\d\rho/\d t=(\nu/2)\Delta\rho$, and finally
$\rho(t;0)=\int d\mu_W^\nu$ which defines a coordinate-free Wiener measure.
Next we let $\phi:\,M\ra\bf C$, ${\cal K}:\,M\times M\ra\bf C$,
$\phi\in({\cal K})L^2(M,\omega)$, and $\phi={\cal K}\phi$ (N.B. $\cal K$ is
an analog of ``polarization''). Quantum dynamics comes from $i\hbar\d\phi/\d
t={\cal H}\phi$, where ${\cal H}={\cal K}h{\cal K}$ (N.B. this relation has
the effect of ``preserving polarization''); also we introduce
$K(T;0):\,M\times M\ra\bf C$, so that $\phi(T)=K(T;0)\phi(0)$. The
construction of the reproducing kernel $\cal K$ and the propagator $K$ reads
  \b  {\cal K}\!\!&\equiv&\!\!\lim_{\nu\ra\infty}2\pi\hbar\, e^{\hbar\nu
T/2}\int e^{(i/\hbar)\int(\theta+dG)}\,d\mu_W^\nu=\lim_{T\ra0}K(T;0)\;,\\
  K(T;0)\!\!&\equiv&\!\!\lim_{\nu\ra\infty}2\pi\hbar\, e^{\hbar\nu T/2}\int
e^{(i/\hbar)\int(\theta+dG-h\,dt)}\,d\mu_W^\nu\;.  \e
Observe again how a flat metric has been used for the Brownian motion; in
our view it is this flat (phase) space that underlies Dirac's remark related
to canonical quantization quoted above.
\subsubsection*{Alternative continuous-time regularizations}
Our introduction of Brownian motion on a {\it flat} two-dimensional phase
space has led to canonical quantization, namely one involving the Heisenberg
operators $P$ and $Q$. If instead we choose to regularize on a phase space
taken as a two-dimensional {\it spherical} surface of radius $R$, where
$R^2\equiv s=\hbar/2,\;\hbar,\;3\hbar/2,\ldots$, then such a Brownian motion
regularization leads to a quantization in which the kinematical operators
are the spin operators $S_1,\;S_2,$ and $S_3$ such that $\Sigma
S_j^2=s(s+1)\hbar^2$, i.e., the generators of the SU(2) group \cite{dkl}. In
like manner, if we introduce a Brownian motion regularization on a
two-dimensional {\it pseudo-sphere} of constant negative curvature, then the
kinematical operators that emerge are the generators of the affine
(``$ax+b$'') group, a subgroup of SU(1,1)\cite{dkp}. The three examples
given here exhaust the simply connected spaces of constant curvature in two
dimensions; they also have the property that the metric assumed for the
Brownian motion regularization coincides with the metric that follows from
the so-derived coherent states.

Summarizing, {\it the geometry of the regularization that supports the
Brownian motion actually determines the nature of the kinematical operators
in the quantization!}
\section*{Regularization on a General 2-D Surface}
Finally, we note that the present kind of quantization can be extended to a
general two-dimensional surface without symmetry and with an arbitrary
number of handles. We only quote the result for the propagator. Let
$\xi^j,\;j=1,2,$ denote the two coordinates, $g_{jk}(\xi)$ the metric,
$a_j(\xi)$ a two-vector, $f_{jk}(\xi)=\d_ja_k(\xi)-\d_ka_j(\xi)$ its curl,
and $h(\xi)$ the classical Hamiltonian. Then the propagator is defined by
\cite{akl}
 \b  \<\xi'',T|\xi',0\>\!\!&=&\!\!\<\xi''|\,e^{-i{\cal H}T/\hbar}|\xi'\>\\
&=&\!\!\lim_{\nu\ra\infty}{\cal
M}_\nu\int\exp\{(i/\hbar)\int[a_j(\xi){\dot\xi}^j-h(\xi)]\,dt\}\\
&&\times\exp\{-(1/2\nu)\int g_{jk}(\xi){\dot\xi}^j{\dot\xi}^k\,dt\}\\
&&\times\exp\{(\hbar\nu/4)\int\sqrt{g(\xi)}\epsilon^{jk}f_{jk}(\xi)\,dt\}\,\
Pi_t\sqrt{g(\xi)}\,d\xi^1\,d\xi^2\;.  \e
Observe, in this general setting, that the phase-space metric tensor
$g_{jk}(\xi)$ is one of the necessary {\it inputs} to the process of
quantization under discussion. From this viewpoint the phase-space metric
induced by the coherent states is regarded as a derived quantity, and in the
general situation the two metrics may well differ.
For a compact manifold $M$ it is necessary that $\int
f_{jk}(\xi)d\xi^j\wedge d\xi^k=4\pi\hbar n$, $n=1,2,3,\ldots\,$. In this
case the Hilbert space dimension $D=n+1-{\o g}$, where ${\o g}$ is the
number of handles in the space (genus). Here the states $|\xi\>$ are
coherent states that satisfy
\b  \one\!\!&=&\!\!\int|\xi\>\<\xi|\,\sqrt{g(\xi)}\,d\xi^1\,d\xi^2\;,\\
   {\cal H}\!\!&=&\!\!\int
h(\xi)\,|\xi\>\<\xi|\,\sqrt{g(\xi)}\,d\xi^1\,d\xi^2\;.  \e
Observe that although the states $|\xi\>$ are coherent states in the sense
of this article, there is generally {\it no transitive group} with which
they may be defined. The propagator expression above is manifestly covariant
under arbitrary coordinate transformations, and a gauge transformation of
the vector $a$ introduces a gauge-like contribution that does not appear in
the field $f$. Finally---and contrary to general wisdom---we note that the
weighting in the case of a general geometry is {\it nonuniform} in the sense
that the symplectic form $\omega=f_{jk}\,d\xi^j\wedge d\xi^k/2$ is generally
{\it not} proportional to the volume element $\sqrt{g(\xi)}\,d\xi^1\,d\xi^2$
needed in the resolution of unity and hence in the path integral
construction.
\section*{Acknowledgements}
Thanks are expressed to G.G. Emch and P.L. Robinson for their comments on
the manuscript.

\end{document}